\documentclass[preprint,aps]{revtex4}
\usepackage[dvips]{graphicx}
\def\singlespace {\smallskipamount=3.75pt plus1pt minus1pt
                  \medskipamount=7.5pt plus2pt minus2pt
                  \bigskipamount=15pt plus4pt minus4pt
                  \normalbaselineskip=12pt plus0pt minus0pt
                  \normallineskip=1pt
                  \normallineskiplimit=0pt
                  \jot=3.75pt
                  {\def\smallskip {\vskip\smallskipamount}}
                  {\def\medskip   {\vskip\medskipamount}}
                  {\def\bigskip   {\vskip\bigskipamount}}
                  {\setbox\strutbox=\hbox{\vrule
                    height10.5pt depth4.5pt width 0pt}}
                  \parskip 7.5pt
                  \normalbaselines}
\def\middlespace {\smallskipamount=5.625pt plus1.5pt minus1.5pt
                  \medskipamount=11.25pt plus3pt minus3pt
                  \bigskipamount=22.5pt plus6pt minus6pt
                  \normalbaselineskip=22.5pt plus0pt minus0pt
                  \normallineskip=1pt
                  \normallineskiplimit=0pt
                  \jot=5.625pt
                  {\def\smallskip {\vskip\smallskipamount}}
                  {\def\medskip   {\vskip\medskipamount}}
                  {\def\bigskip   {\vskip\bigskipamount}}
                  {\setbox\strutbox=\hbox{\vrule
                    height15.75pt depth6.75pt width 0pt}}
                  \parskip 11.25pt
                  \normalbaselines}
\def\doublespace {\smallskipamount=7.5pt plus2pt minus2pt
                  \medskipamount=15pt plus4pt minus4pt
                  \bigskipamount=30pt plus8pt minus8pt
                  \normalbaselineskip=30pt plus0pt minus0pt
                  \normallineskip=2pt
                  \normallineskiplimit=0pt
                  \jot=7.5pt
                  {\def\smallskip {\vskip\smallskipamount}}
                  {\def\medskip   {\vskip\medskipamount}}
                  {\def\bigskip   {\vskip\bigskipamount}}
                  {\setbox\strutbox=\hbox{\vrule
                    height21.0pt depth9.0pt width 0pt}}
                  \parskip 15.0pt
                  \normalbaselines}

\newcommand{\beq}{\begin{equation}}
\newcommand{\eeq}{\end{equation}}
\newcommand{\bea}{\begin{eqnarray}}
\newcommand{\eea}{\end{eqnarray}}

\begin{document}

\preprint{
\hfill$\vcenter{\hbox{\bf IUHET-521}}$  }
\title{\vspace*{.75in}
Wave Packets in Discrete Quantum Phase Space}

\author{Jang Young Bang\footnote{Electronic address:berger@indiana.edu}
 and Micheal S. Berger\footnote{Electronic address:jybang@indiana.edu}}

\affiliation{Physics Department, Indiana University, Bloomington, IN 47405, USA}

\date{\today}

\begin{abstract}
The properties of quantum mechanics with a discrete phase space are studied.
The minimum uncertainty states are found, and these states
become the Gaussian wave packets
in the continuum limit. With a suitably chosen
Hamiltonian that gives free particle motion in the continuum limit, it is found
that full or approximate periodic time evolution can result. 
This represents an example of revivals 
of wave packets that in the continuum limit is the familiar free 
particle motion on a line. 
Finally we examine the uncertainty principle for discrete phase space and obtain
the correction terms to the continuum case.
\end{abstract}

\maketitle

\section{Introduction}

There are fundamental differences between quantum mechanics on a continuous 
configuration space with that on a discrete one. In recent years it has 
become fashionable to consider the possibility that spacetime is discrete and
the continuous space time we experience is an emergent property that is evident
only at long distances. In particular the discreteness is suggested by quantum
mechanics, while classical gravity and continuous spacetime are described 
by general relativity. Therefore a discrete spacetime is natural concept
that arises naturally in 
attempts to combine quantum mechanics and general relativity. 

In Ref.~\cite{Bang:2006va} the authors presented a simple model in which 
momentum was compactified and the corresponding position operator has discrete
eigenvalues. The continuum limit corresponds to the familiar quantum mechanics
on a line. In the context of the model it was shown that the 
uncertainty principle 
obtains corrections of the form of a generalized uncertainty principle 
(GUP)\cite{Konishi:1989wk}.
The model provided a way to see explicitly corrections that are 
naturally present in a quantum theory with a discrete configuration space.
It must be considered a toy model as it makes no attempt to incorporate gravity.
Further, since the continuum theory can be the limit of 
different discrete theories, 
the results presented there 
should be only considered as representative of the type of
corrections that can be obtained. 

In addition to the possible applications in describing a quantum spacetime, 
discrete quantum systems are interesting 
in their own right.
The concept of considering compactified momentum has been considered
briefly before in the literature. Schwinger\cite{Schwinger} introduced
a complete basis of unitary operators that form a realization of the
Heisenberg group.  The discrete quantum phase space was examined in a
series of subsequent papers by Santhanam and
collaborators\cite{Santhanam:1978ka}
and Galetti and
collaborators\cite{Galetti:1994zh}.
In
these later papers, the connection with the continuum limit was
investigated and it was shown how the usual case of quantum mechanics on a 
line can
be obtained by taking the lattice spacing to zero (in both position
and momentum space). 

In this paper we consider dynamics in a discrete quantum phase space. 
Our interest is twofold: First we show how one can define a 
Hamiltonian on the discrete quantum phase space which has the appropriate 
classical limit of a free particle. A Hamiltonian allows one to study time 
evolution and therefore possible dynamics. Here this is naturally 
achieved in the free
particle motion where phase space is a discrete torus.
We are able to show that a suitably defined Hamiltonian 
gives time evolution that is in some cases exactly periodic. 
This behavior is reminiscent of the 
revival of Rydberg states\cite{Parker:1986zz,Alber:1986zz,Bluhm:1995nr}. 
In the discrete
quantum phase space considered here there is no potential, and the revivals
occur for a freely moving wave packet. For small times
and sufficiently fine discretization of the phase space, the motion of the 
wave packet will appear to spread like the Gaussian wave packet on the line. 
For some cases the revivals are approximate, and occur only for states that 
are minimum uncertainty states. Our second result  
continues the direction of Ref.~\cite{Bang:2006va}, where a full quantum 
mechanical model with a discrete configuration was considered. In that paper
the momentum was compactified on a circle leaving a discrete spectrum for 
the position (phase) operator. It was shown that the uncertainty principle 
involving the position and momentum operators receives the expected corrections
that have been postulated in quantum gravity theories including string theory.
The modified uncertainty principle has been called the generalized uncertainty 
principle (GUP) in this context.

\section{Minimum Uncertainty State}

Consider a quantum system for which both the position operator $\hat{U}$ 
and the 
momentum operator $\hat{V}$ have discrete eigenvalues. The most straightforward
example case occurs when both 
are compactified on circles and are related by a discrete Fourier transform.
This may be called a discrete torus, and the
 model was first considered by Schwinger\cite{Schwinger}.
It differs from the model(s) in Ref.~\cite{Bang:2006va} where only one variable
(space or momentum) was discretized at a time.

To make contact with the continuum limit, we relate these 
unitary operators to
the usual Hermitian position $\hat{P}$ and momentum $\hat{Q}$ operators. 
Our phase 
space is a torus so we define
\bea
\hat{V}&=&\exp[i\beta\hat{P}]\\
\hat{U}&=&\exp[i\alpha\hat{Q}]\;,
\eea
where
\bea
\alpha\beta&=&{{2\pi}\over {\hbar N}}\;.
\label{alphabeta}
\eea
It is well-known that the appropriate operators to use for quantum mechanics
on a circle should be unitary ones like $\hat{U}$ and $\hat{V}$ 
and can be referred to 
as phase operators. 
In the limit $N\to \infty$ one imagines that the solutions approaches the 
continuum limit, 
which is simply the quantum mechanics of a particle on the line. 
The usual position and momentum operators, $\hat{Q}$ and 
$\hat{P}$, recover their usual meaning in the continuum limit.
 For finite $N$ one can attempt to calculate the $1/N$ 
corrections that arise from the small (for that case) level of discreteness
that persists. 
The commutation relations are
\bea
\hat{V}\hat{U}&=&\exp[2\pi i/N]\hat{U}\hat{V}\;.
\eea
This algebra arises in the 
study of confinement and is sometimes also called 
the 't~Hooft algebra\cite{'tHooft:1977hy}. It also arises in the matrix theory 
approach to string theory. See, for example, Ref.~\cite{Taylor:2001vb}. 
It is also of interest in studies of 
the noncommutative torus\cite{Connes:1997cr}.
On the eigenstates $|U_j\rangle$ of the $U$ operator
\bea
\hat{V}\left|u_j\right\rangle&=&\left|u_{j-1}\right\rangle,\\
\hat{U}\left|u_j\right\rangle&=&u_j\left|u_{j}\right\rangle,\;
\eea
where the eigenvalue is the phase 
\bea
u_j&=&\exp[2\pi i j/N].\;
\eea
So $\hat{V}$ and $\hat{V}^\dagger$ are the 
lowering and raising operators respectively
(this choice is 
conventional) when acting on 
the discrete eigenstates of position.

We now proceed to discuss the minimum uncertainty states (MUS) for this quantum
system. By minimum uncertainty states we mean states that saturate the 
uncertainty principle. In the continuum limit these states approach the 
familiar
Gaussian states with minimum uncertainty. These states can be derived by a 
straightforward generalization of the standard argument for Hermitian
operators. 

A minimum uncertainty state with respect to the operators $\hat{U}$ and 
$\hat{V}$ must satisfy the following
\cite{Bang:2006va,Tanimura:1993hf} \bea \left
  (\hat{V}-\left\langle\hat{V}\right\rangle\right )\left
  |\psi\right\rangle &=&-\lambda\left (\hat{U}-\left\langle
    \hat{U}\right\rangle\right )\left |\psi\right\rangle.\; \eea This
can be expressed as follows

\bea \hat{M}\left|\psi\right\rangle &=&\mu\left|\psi\right\rangle\;
\eea where $\hat{M}=\hat{V}+\lambda\hat{U}$, $\mu=\left\langle
  \hat{V}\right\rangle+\lambda\left\langle \hat{U}\right\rangle$ and  
we define coefficients $c_j$ so that
$\left|\psi\right\rangle=\sum_{j=0}^{N-1}c_j \left|u_j\right\rangle$.
Solving this yields \bea {c_{j+1}}\over{c_j}&=&\Delta_j,\; \eea where
\bea
\Delta_j=\mu-\lambda
\exp[2\pi i j/N].\; \eea The periodic condition ($c_0=c_N$) requires
\bea \prod_{k=0}^{N-1}\Delta_k&=&1.\; \eea This condition determines $\lambda$,
in terms of $\mu$, or equivalently in terms of $\langle\hat{U}\rangle$ 
and $\langle\hat{V}\rangle.$ It reads
\bea
\mu^N-\lambda^N=1\;.
\eea
The coefficients $c_j$ are then determined up to normalization
(The unnormalized coefficients are simply $c_j=\prod_{k=0}^{j-1}\Delta_k$).
One can establish a momentum space wave function
\bea
\left|\psi\right\rangle=\sum_{k=0}^{N-1}d_k \left|v_k\right\rangle\;,
\eea
where the coefficients are determined from the $c_j$ by a discrete Fourier
transform. A simple case of interest is the state $|\psi\rangle=|u_j\rangle$
for some $j$. This state is an eigenstate of the position operator $\hat{U}$,
and its possible momentum space wave functions are characterized 
by $|d_k|^2=1/N$. There is no dispersion in $\hat{U}$, while the dispersion in 
$\hat{V}$ is finite, since there are only a finite number $N$ of sites 
$|v_k\rangle$.  

\begin{figure}
        \centering
                \includegraphics{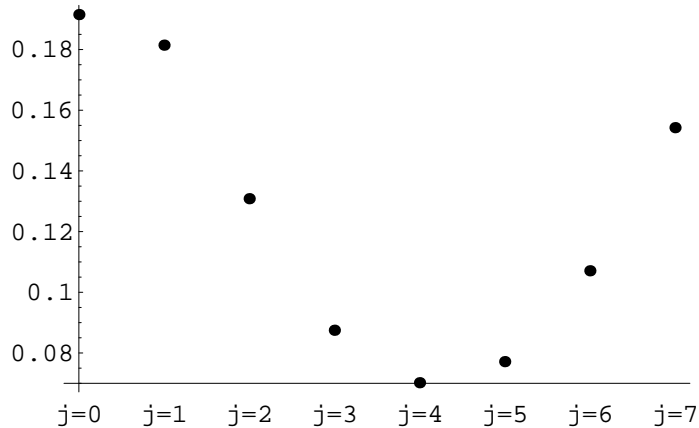}
        \caption{The coefficients $|c_j|^2$ for 
N = 8 and $\langle V\rangle = \langle U\rangle =1/2$. This wave packet 
has approximately a (discrete) Gaussian profile.}
        \label{fig:cj}
\end{figure}

More generally the minimum uncertainty states
become approximately Gaussians in the continuum limit. 
An illustrative example is 
shown in Fig.~\ref{fig:cj} for the case of $N=8$ for 
$\langle \hat{V}\rangle = \langle \hat{U}\rangle =1/2$. 
An important point is that the expectation values of $\hat{U}$ and $\hat{V}$ 
for a wave packet do not need to lie on
the points of either the position or momentum circles (but, as an average, must
lie within each circle). The shape is similar to the 
Gaussian for the continuous case and exhibits the mod $N$ behavior necessary 
for a properly defined wave packet. In fact one can show that as the the 
discretization becomes increasingly fine, the minimum uncertainty state 
approaches the familiar Gaussian minimum uncertainty wave packet for quantum 
mechanics on the line. The momentum space wave function also approaches the 
Gaussian shape in the continuum limit. Choosing a discretization for both 
the position and momentum spaces allows us to maintain a symmetry between the 
two.

\section{Time Evolution}

For quantum mechanics on a line, the minimum uncertainty states (Gaussians in 
both position and momentum spaces) do not continue to saturate the 
uncertainty relation under the time evolution 
dictated by the free particle Hamiltonian. The spreading of the wave packet in
the position representation is easy to understand as a consequence of the 
nonzero dispersion of the wave in the momentum representation. The various 
momentum components of the position space wave packet move with different 
'velocities' resulting in the increased spread at later times. 

However, placing the
wave packet on the discrete toroidal lattice resulting from
compactifying both coordinate and momentum spaces allows the minimum 
uncertainty wave
packet to disperse, and at some later time arrive again at 
its starting 
position with approximately its original shape. 
We now proceed to define a Hamiltonian on
our discrete space that describes the time evolution. 

Schwinger has shown that one can construct a complete orthogonal base
system using the following set of operators\cite{Schwinger}:
\bea \hat{S}_{mn}&=&e^{(i\pi/N)mn}\hat{U}^m\hat{V}^n.\; \eea Some of
their properties are
\begin{enumerate}
        \item any operator can be written in this base:
          \bea
          \hat{O}&=&\sum_{m,n} O_{mn}\hat{S}_{mn},\nonumber\\
    {\rm where}\:\:  O_{mn}&=&tr[\hat{S}^\dag_{mn} \hat{O}];\;
    \eea
  \item they have the following action on the ket:
    \bea
    \hat{S}_{mn}\left|u_j\right>&=&e^{(i\pi/N)[2j-n]m}\left|u_{j-n}\right>;\label{action}\;
    \eea
  \item their (group) product becomes:
    \bea
    \hat{S}_{rs}\hat{S}_{mn}&=&e^{(i\pi/N)[ms-nr]}\hat{S}_{(m+r)(n+s)};\label{product}\;
    \eea
  \item there is the (group) identity:
    \bea
    \hat{S}_{00}&\equiv&\textbf{1};\;
    \eea
  \item the Hermitian conjugate and the inverse of a given element is:
    \bea
    \hat{S}^\dag_{mn}=\hat{S}^{-1}_{mn}=\hat{S}_{-m-n};\;
    \eea
  \item under a similarity transformation, they become:
    \bea
    \hat{S}_{mn}\hat{S}_{rs}\hat{S}_{mn}^{-1}&=&e^{(2\pi i/N)[2j-n]m}\hat{S}_{rs};\;
    \eea
  \item they satisfy associativity:
    \bea \left(\hat{S}_{mn}\hat{S}_{rs}\right)\hat{S}_{kl}&=&\hat{S}_{mn}\left(\hat{S}_{rs}\hat{S}_{kl}\right).\label{assoc}\;
    \eea
\end{enumerate}

The labels $\left(m,n\right)$ of the operator $\hat{S}_{mn}$ occupy
discrete points on the lattice defined on the toroidal surface and one
can use these labels to describe the quantum phase space.
These operators are constructed in such a way to exhibit features resembling
the symplectic structure of classical mechanics\cite{Aldrovandi:1990wy}. For
our purposes,
is not crucial to understand these properties of the operators. However it is
necessary to define a new representation of them for which the overall phase
depends on the state on which the operator acts.  

Define new operators 
\bea
\hat{T}^j_{mn}&\equiv&e^{-i\alpha_1(j;(m,n))}\hat{S}_{mn}\;,
 \eea
where
\bea
\alpha_1(j;(m,n))&=&(\pi/N)[2j-m]n\;.  \eea
This is something of an abuse of notation since, as stated above, 
the phase of the operator 
depends on the state $|u_j\rangle$ on which it is acting. The detailed 
description of the representation theory for the operators can be found in 
Ref.~\cite{Aldrovandi:1990wy}.
From our perspective the purpose of the extra phase in the operator simply 
multiplies the primitive $\hat{S}$ operators by the appropriate phase as one
moves around the discretized torus.

For the purpose of 
proposing a time evolution for the wave packet, we assume here that time
can be taken to be a continuous variable and propose a hermitian Hamiltonian 
so that the evolution is unitary. (One can alternatively assume that time 
evolution is given by a discrete shift operator, say the operator $\hat{V}$, 
that 
trivially gives a free particle motion for any wave function. This does not 
approach the usual continuum limit where a wave packet is expected to spread
because of its distribution in momenta.) To maintain the symmetry between
configuration space and momentum space, the Hamiltonian should involve 
those
operators $\hat{T}^j_{mn}$ with $m+n=N$, or equivalently $m=k,n=-k$,
where $1\leq k \leq N-1$.  
From the (unitary) operator $\hat{T}^j_{k,-k}$ and its conjugate, one can 
construct a Hermitian Hamiltonian. 
For $k\ll N$ it will 
also yield time evolution that gives the usual free particle motion in the 
continuum limit.

Consider then the
following (dimensionless) Hamiltonian: 
\bea
\hat{H}&=&2-\hat{T}^j_{k,-k}-\hat{T}^{j\dag}_{k,-k}\;.  
\eea
Our convention will be to take time to be dimensionless, but one can put in 
a scale $t_0$ that is otherwise undetermined.
Notice also that the operator $\hat{T}^j_{k,-k}$ will generate the same amount
(assuming $k\ll N$) of translation in both configuration and momentum
spaces.  

Consider a wave packet localized near $j=0$.
In the continuum limit, $\hat{H}$ takes the following form: 
\bea
\hat H &\approx&k^2\beta^2\left[\hat P^2+{\alpha^2\over\beta^2}\hat Q^2-{\alpha\over\beta}\left\{\hat Q,\hat P \right\}-\left(\alpha k\hbar-{4\pi j\over{\beta N}} \right)\hat P\right.\nonumber\\
&&+\left.\left({k\hbar\alpha^2\over\beta}-{4\pi j\alpha\over{\beta^2 N}} \right)\hat Q+\ldots \right]\nonumber \\
&=&k^2\beta^2\left[\hat P^2+{\alpha^2\over\beta^2}\hat Q^2-{\alpha\over\beta}\left\{\hat Q,\hat P \right\}-{1\over \beta}\hat P\left({2\pi\over{N}}[k-2j]\right)\right.\nonumber\\
&&+\left.{\alpha\over\beta^2}\hat Q\left({2\pi\over{N}}[k-2j] \right)+\ldots \right]
\eea
The parameter $k^2$ in the definition can be interpreted as proportional to 
the inverse mass of the wave packet.

Motivated by a potential application to quantum gravity, 
one can  take the discretization scale to be the Planck length $\ell_p$. The 
compactification radius 
$R$ of the configuration space is
\bea
N\ell_p&=&2\pi R\;. \eea 
A specification of $\alpha $ and $\beta$ 
consistent with Eq.~(\ref{alphabeta}) is
\bea
\alpha&=&\sqrt{{{2\pi}\over{N^{3/2}\ell_p^2}}};\;
\beta=\sqrt{{{2\pi\ell_p^2}\over{N^{1/2}\hbar^2}}}\;,  
\label{defnalphabeta}
\eea 
which for large $N$ has the interpretation of a free particle Hamiltonian.

\begin{figure}[t]
\centerline{
\mbox{\includegraphics[width=2.50in]{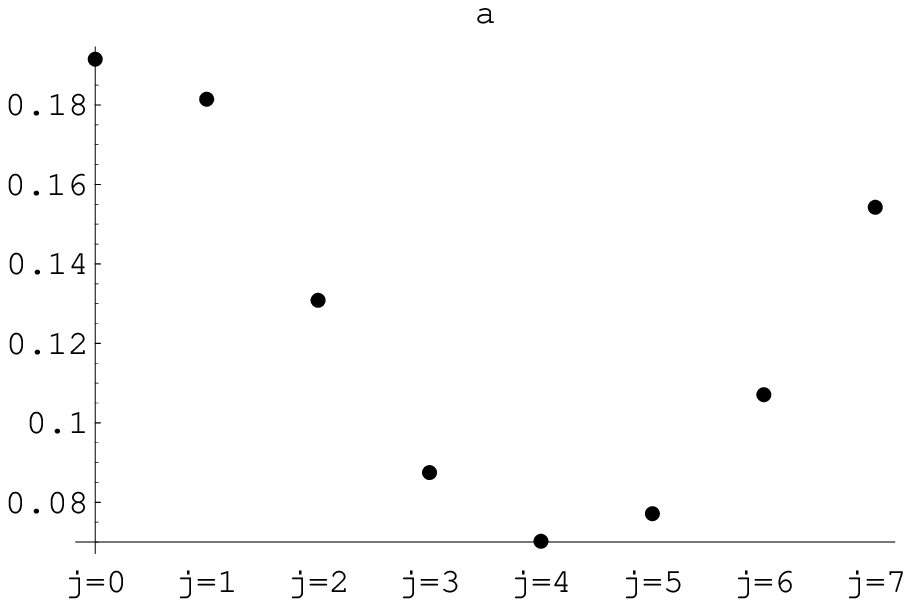}}
\mbox{\includegraphics[width=2.50in]{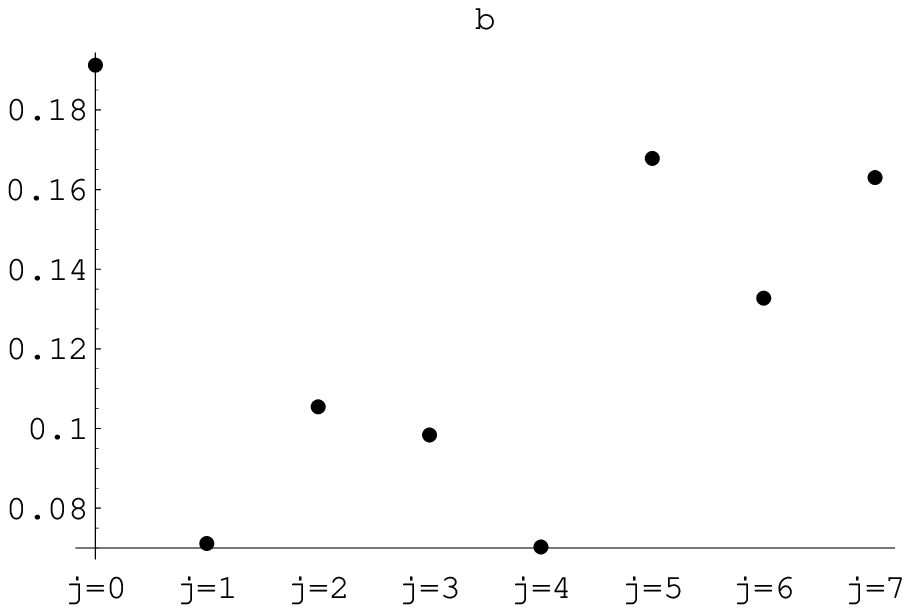}}
}
\vspace*{0.5in}
\centerline{
\mbox{\includegraphics[width=2.50in]{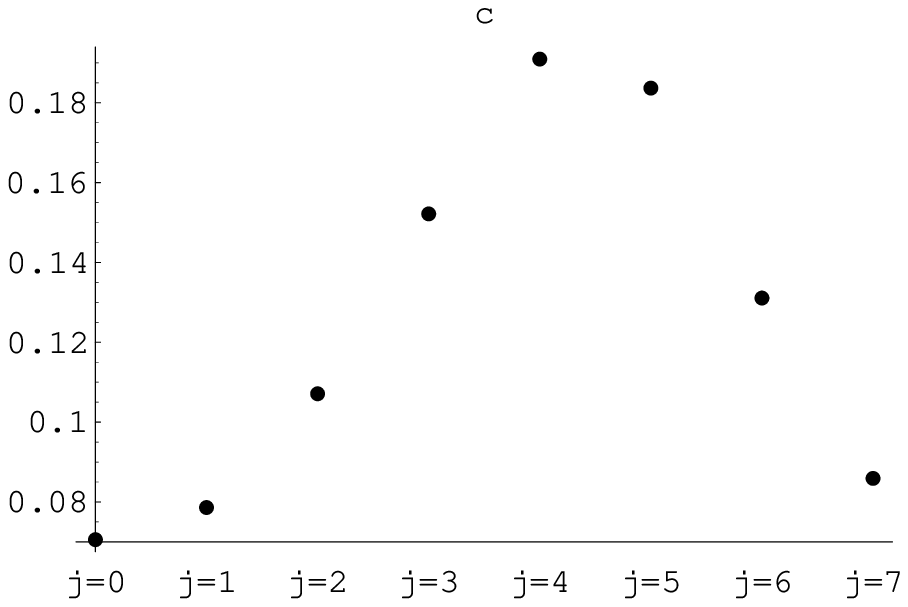}}
\mbox{\includegraphics[width=2.50in]{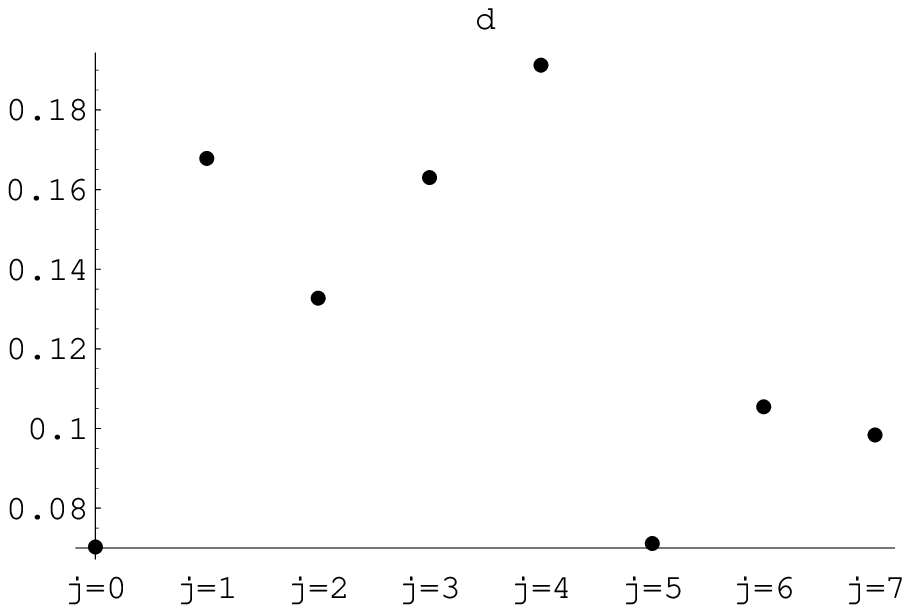}}
}
\vspace*{0.5in}
\centerline{
\mbox{\includegraphics[width=2.50in]{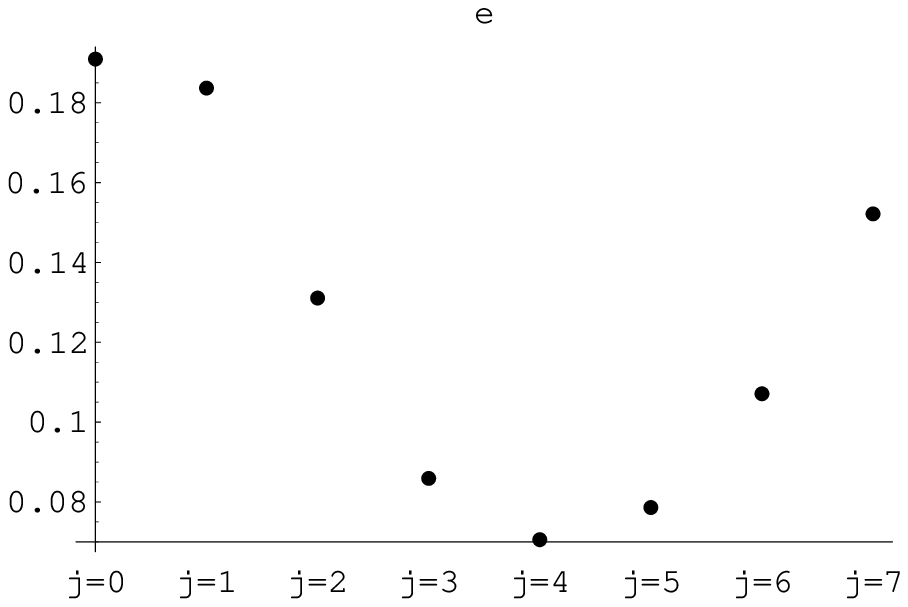}}
}
        \caption{Evolution of a MUS wave packet for $N = 8,k=2$ with $\langle V\rangle = \langle U\rangle =1/2$. The plots correspond to times (a) $t=0$ 
(b) $t=\pi/4$, (c) $t=\pi/2$, (d) $t=3\pi/4$, and (e) $t=\pi$. 
The motion is periodic with 
period $\pi$ and at half-periods the wave packet is located halfway around the 
circle.}
        \label{fig:k2a}
\end{figure}

\begin{figure}[t]
\centerline{
\mbox{\includegraphics[width=2.50in]{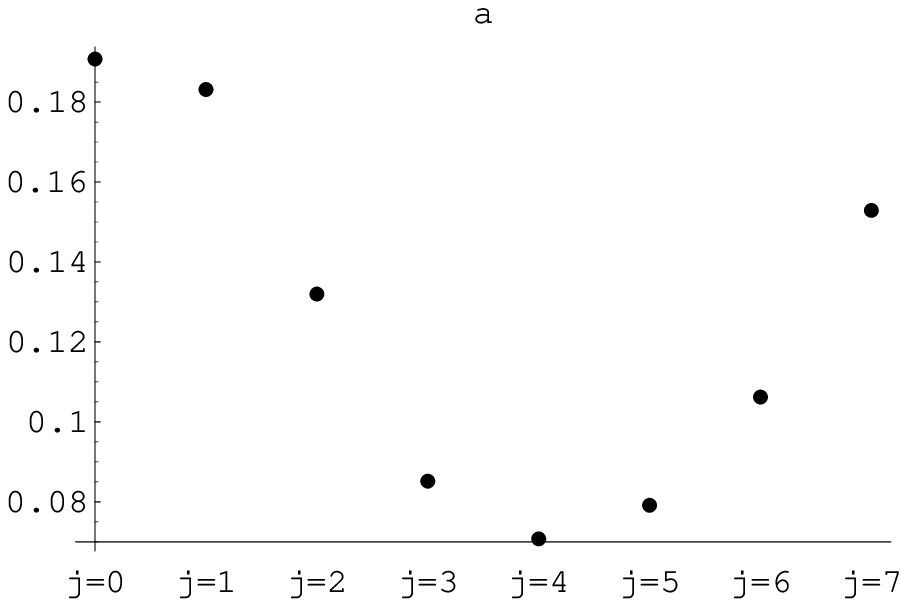}}
\mbox{\includegraphics[width=2.50in]{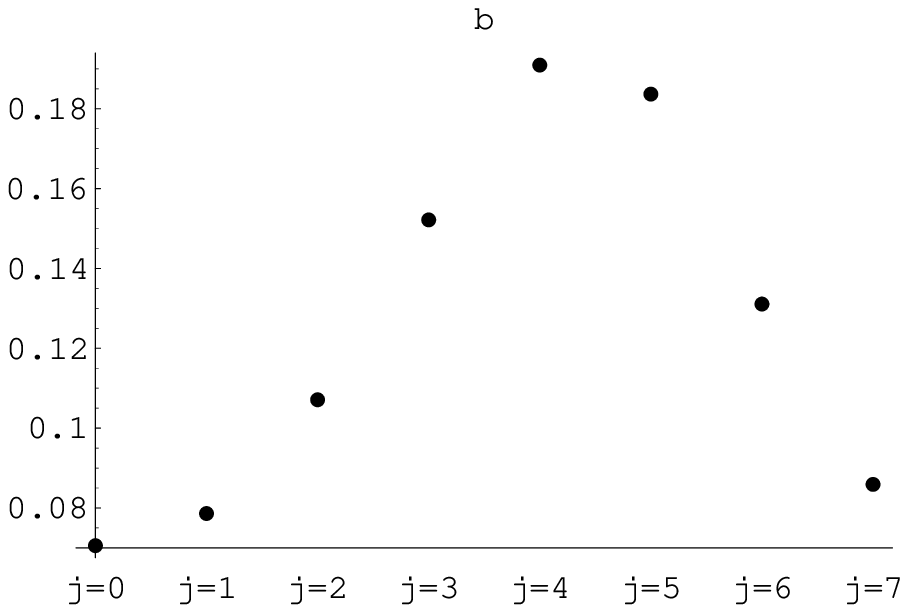}}
}
\vspace*{0.5in}
\centerline{
\mbox{\includegraphics[width=2.50in]{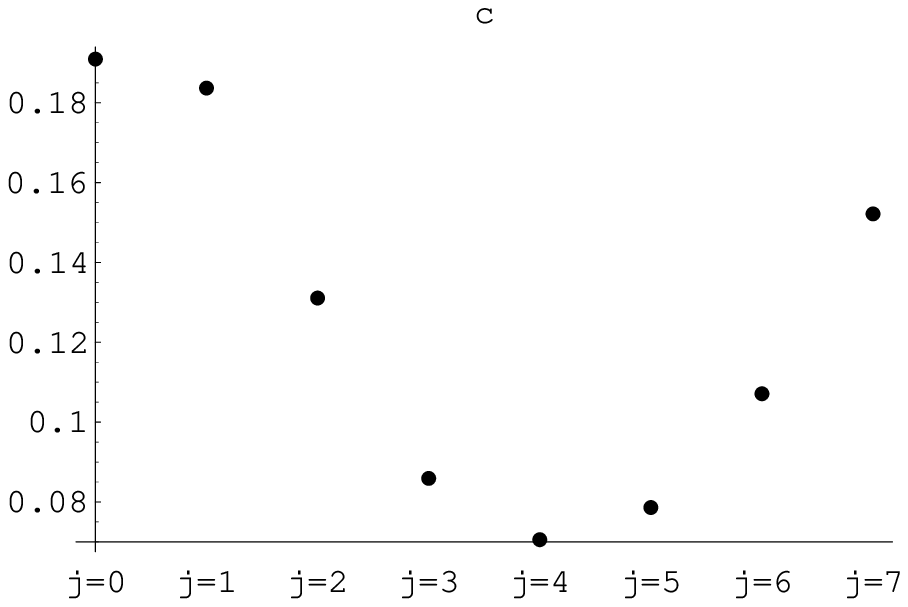}}
\mbox{\includegraphics[width=2.50in]{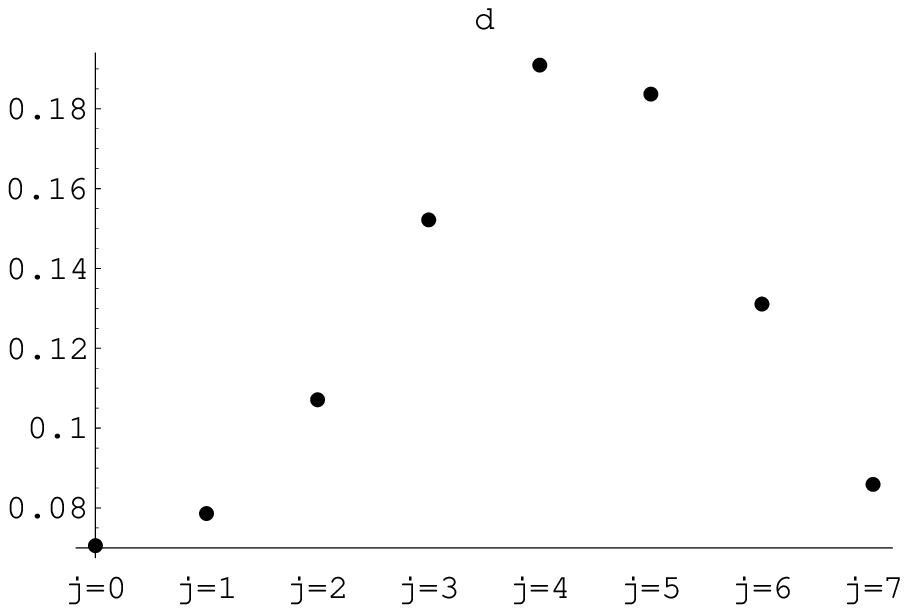}}
}
\vspace*{0.5in}
\centerline{
\mbox{\includegraphics[width=2.50in]{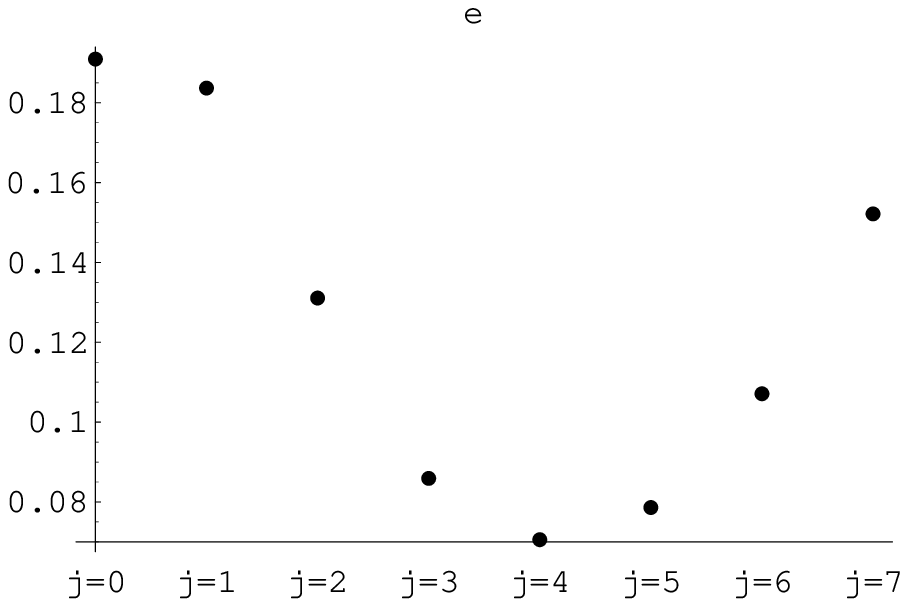}}
}
        \caption{Evolution of a MUS wave packet for $N = 8,k=4$ with 
$\langle V\rangle = \langle U\rangle =1/2$. The plots correspond to times 
(a) $t=0$,
(b) $t=\pi/4$, (c) $t=\pi/2$, (d) $t=3\pi/4$, and (e) $t=\pi$. 
The motion is periodic with 
period $\pi/2$ and 
at half-periods the wave packet is located halfway around the 
circle.}
        \label{fig:k4a}
\end{figure}

\begin{figure}[t]
\centerline{
\mbox{\includegraphics[width=2.50in]{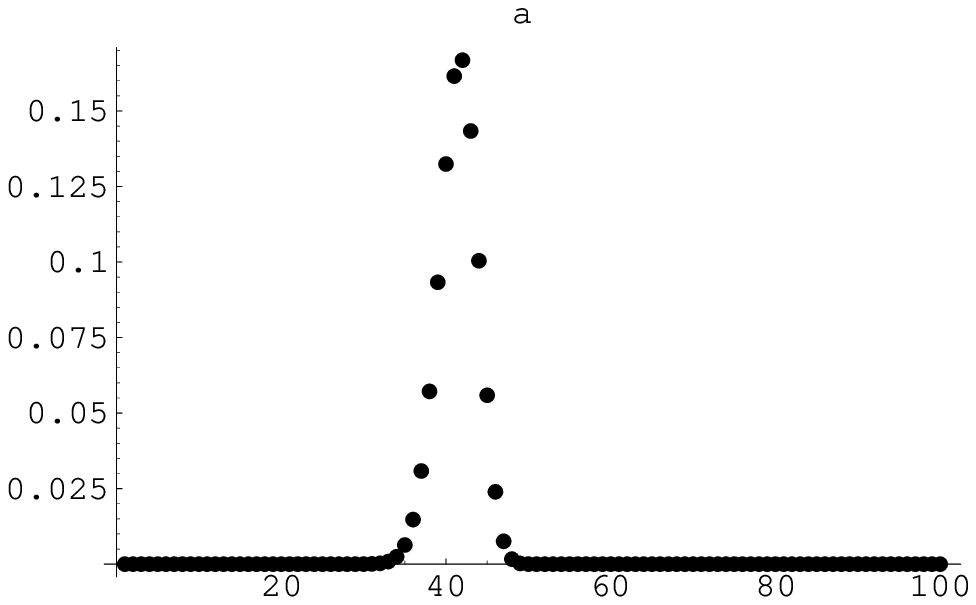}}
\mbox{\includegraphics[width=2.50in]{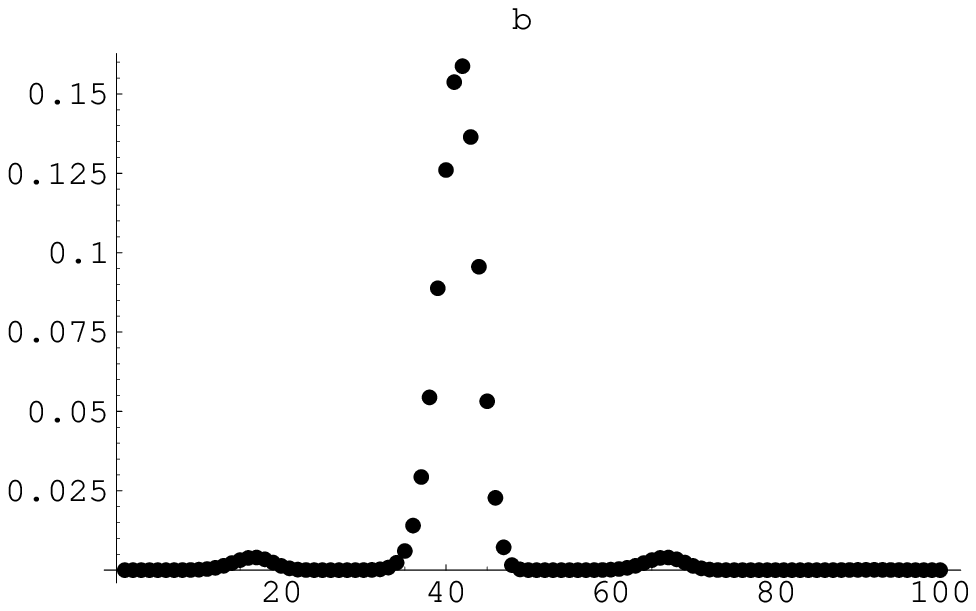}}
}
\vspace*{0.5in}
\centerline{
\mbox{\includegraphics[width=2.50in]{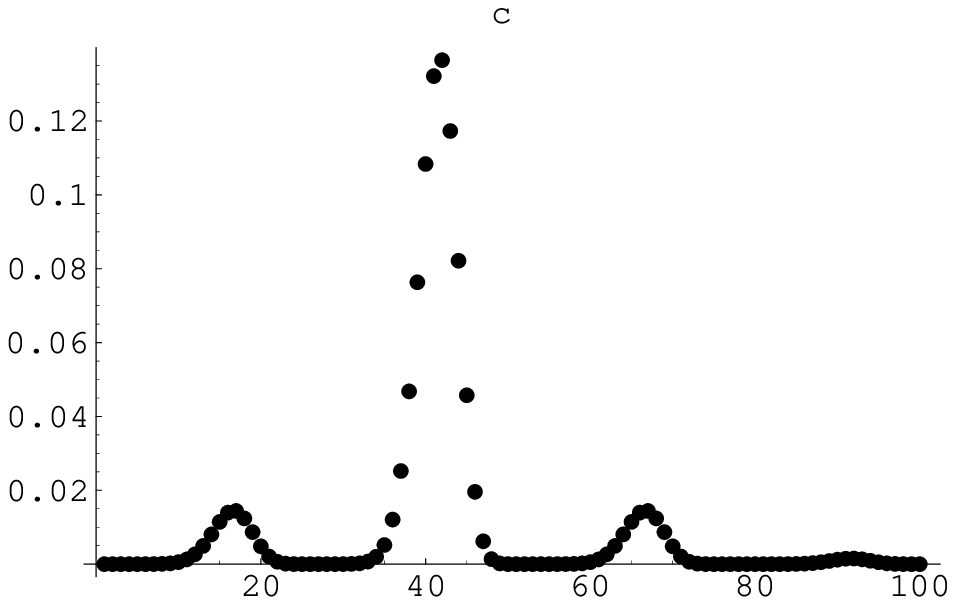}}
\mbox{\includegraphics[width=2.50in]{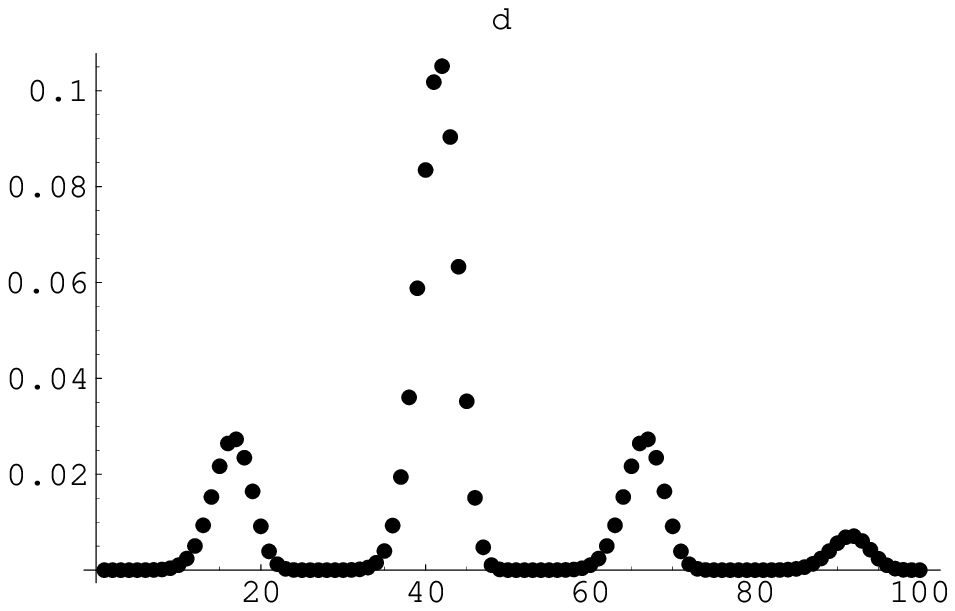}}
}
\vspace*{0.5in}
\centerline{
\mbox{\includegraphics[width=2.50in]{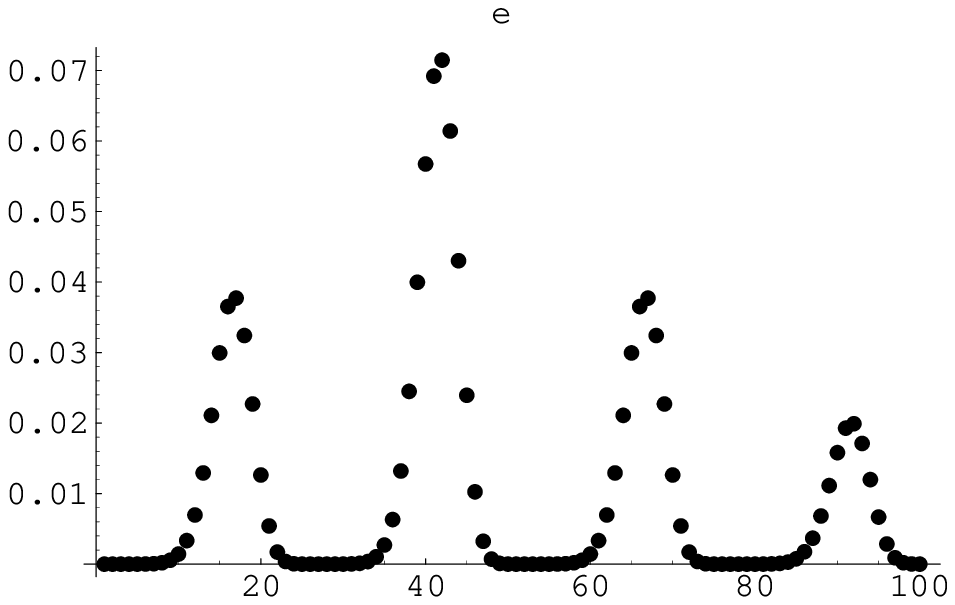}}
\mbox{\includegraphics[width=2.50in]{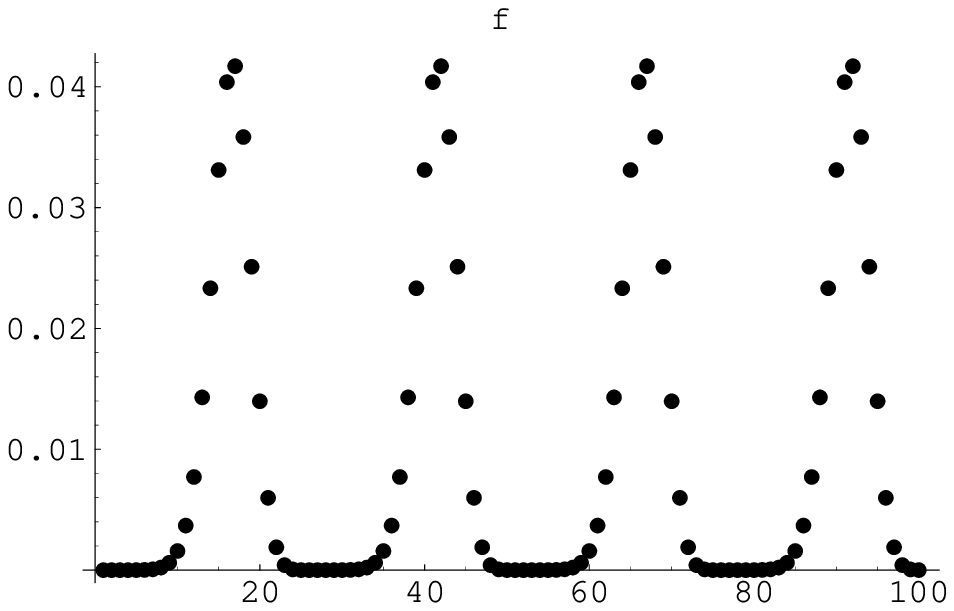}}
}
        \caption{Evolution of a MUS wave packet for $N = 100,k=25$ with 
$\mu=1.5, \lambda=-1.497+0.094i$. The plots correspond to times 
(a) $t=0$, (b) $t=5\pi/100$, (c) $t=10\pi/100$, (d) $t=15\pi/100$,
(e) $t=20\pi/100$, and (f) $t=25\pi/100$. The motion is periodic with 
period $\pi$ and 
at half-periods the wave packet is located halfway around the 
circle.}
        \label{fig:k100a}
\end{figure}

\begin{figure}[t]
\centerline{
\mbox{\includegraphics[width=2.50in]{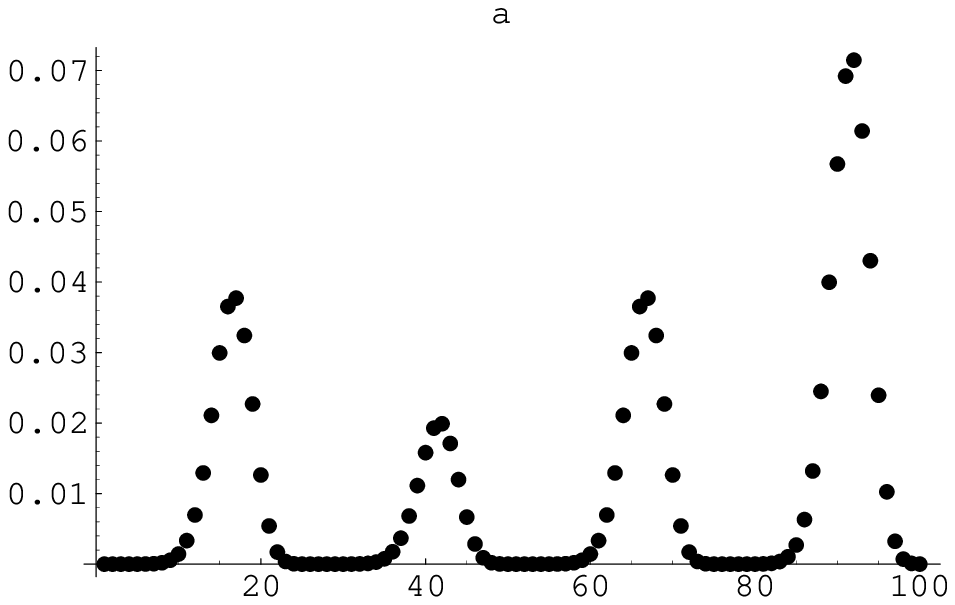}}
\mbox{\includegraphics[width=2.50in]{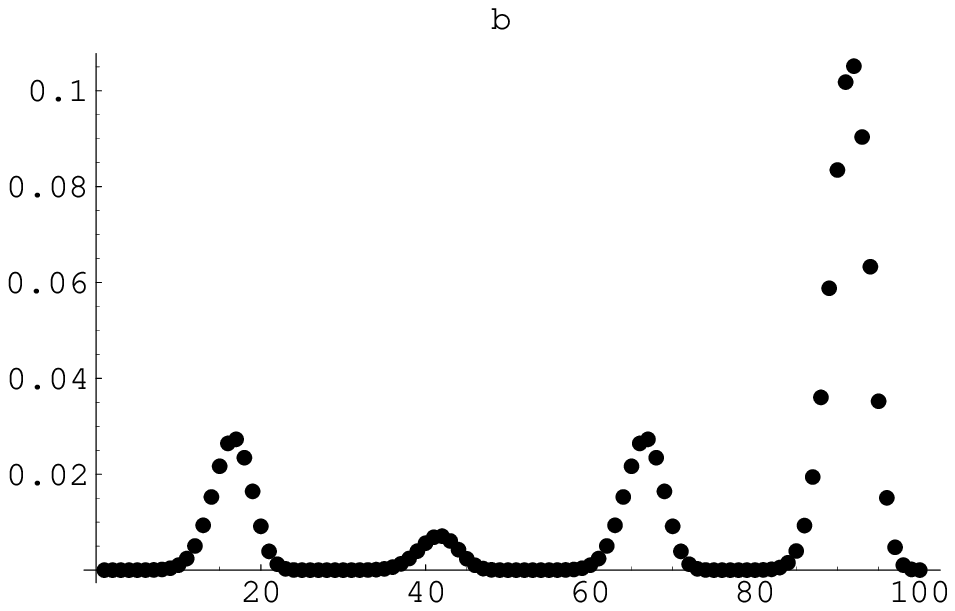}}
}
\vspace*{0.5in}
\centerline{
\mbox{\includegraphics[width=2.50in]{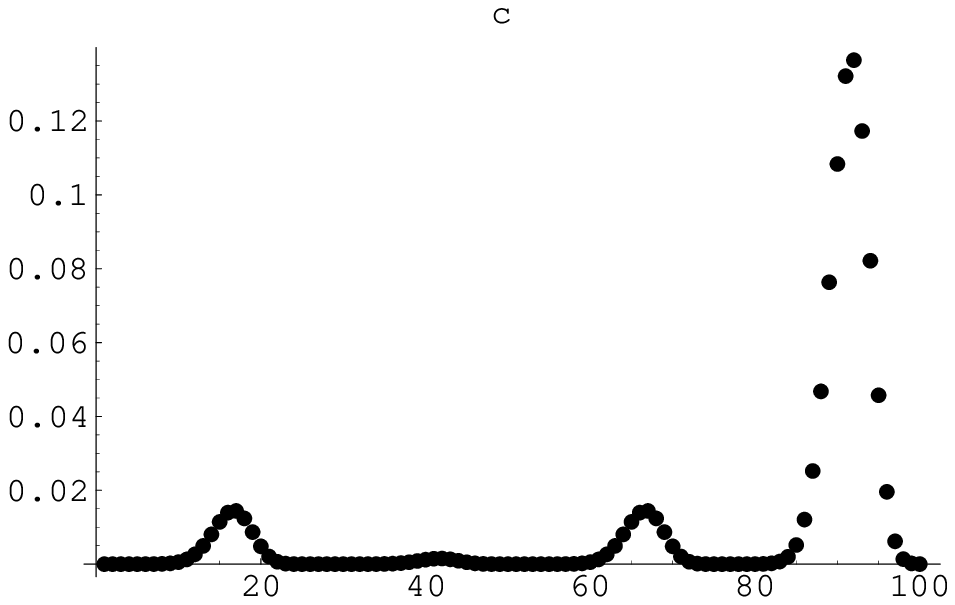}}
\mbox{\includegraphics[width=2.50in]{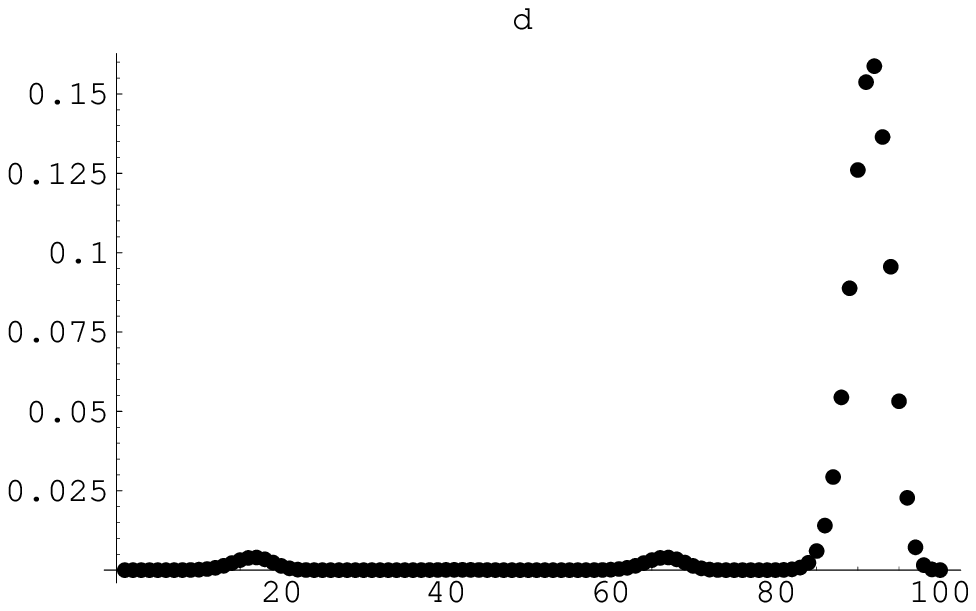}}
}
\vspace*{0.5in}
\centerline{
\mbox{\includegraphics[width=2.50in]{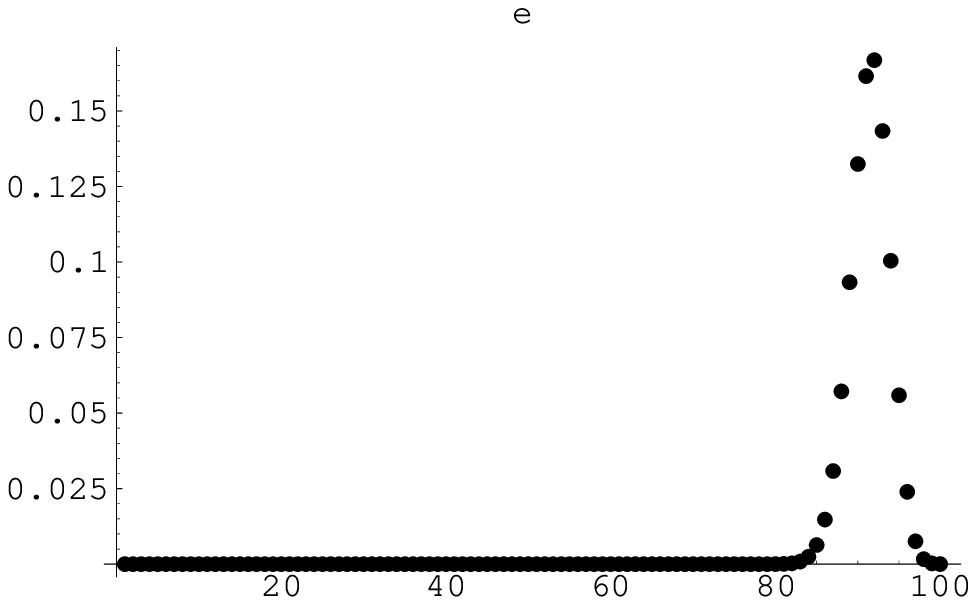}}
}
        \caption{The continuation of the evolution of a MUS wave packet for 
$N = 100,k=25$ in Fig.~\ref{fig:k100a} with $\mu=1.5, \lambda=-1.497+0.094i$. 
The plots correspond to times (a) $t=30\pi/100$,
(b) $t=35\pi/100$, (c) $t=40\pi/100$, (d) $t=45\pi/00$, and 
(e) $t=50\pi/100$. The motion is periodic with 
period $\pi$ and 
at half-periods the wave packet is located halfway around the 
circle.}
        \label{fig:k100b}
\end{figure}

\begin{figure}[t]
\centerline{
\mbox{\includegraphics[width=2.50in]{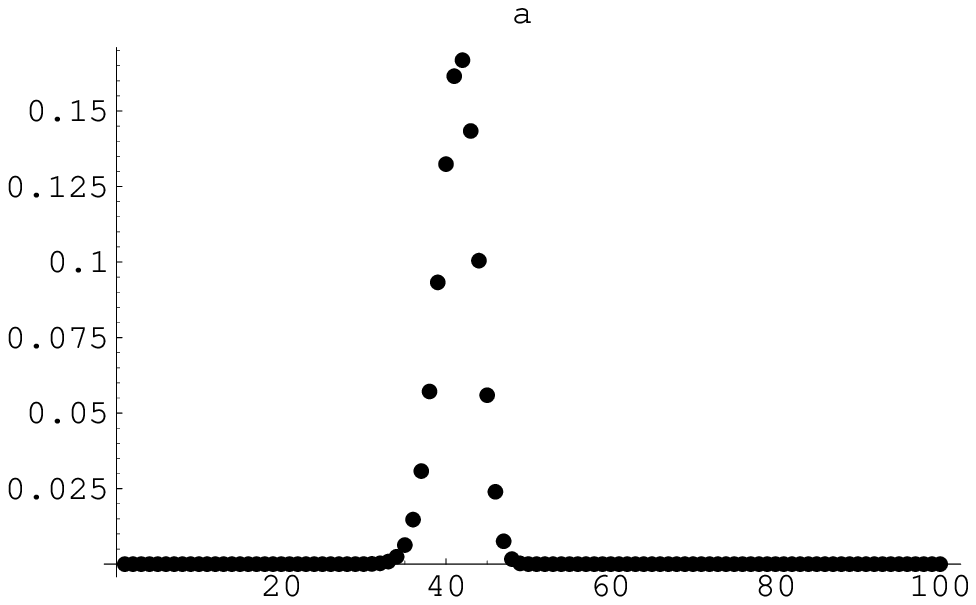}}
\mbox{\includegraphics[width=2.50in]{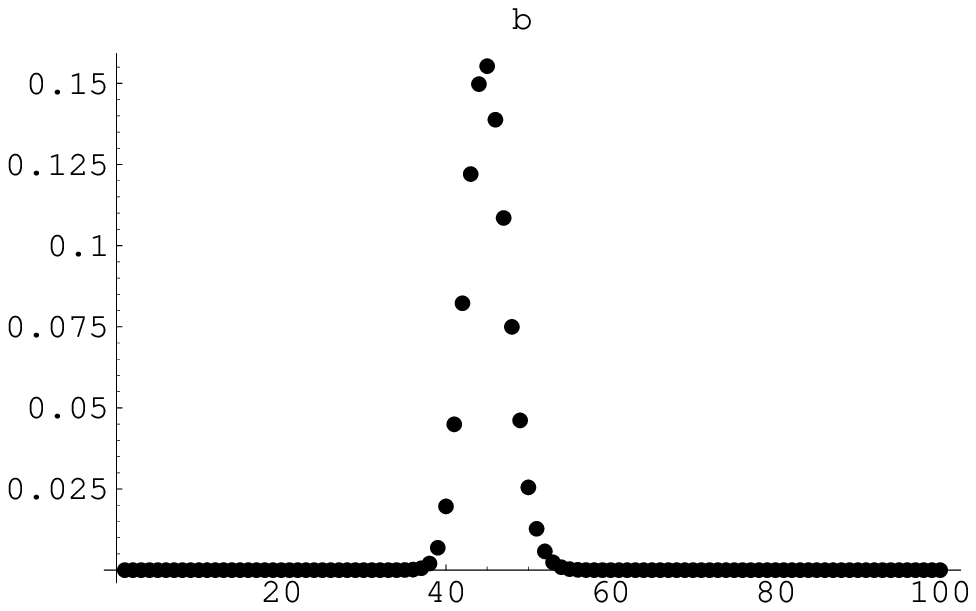}}
}
\vspace*{0.5in}
\centerline{
\mbox{\includegraphics[width=2.50in]{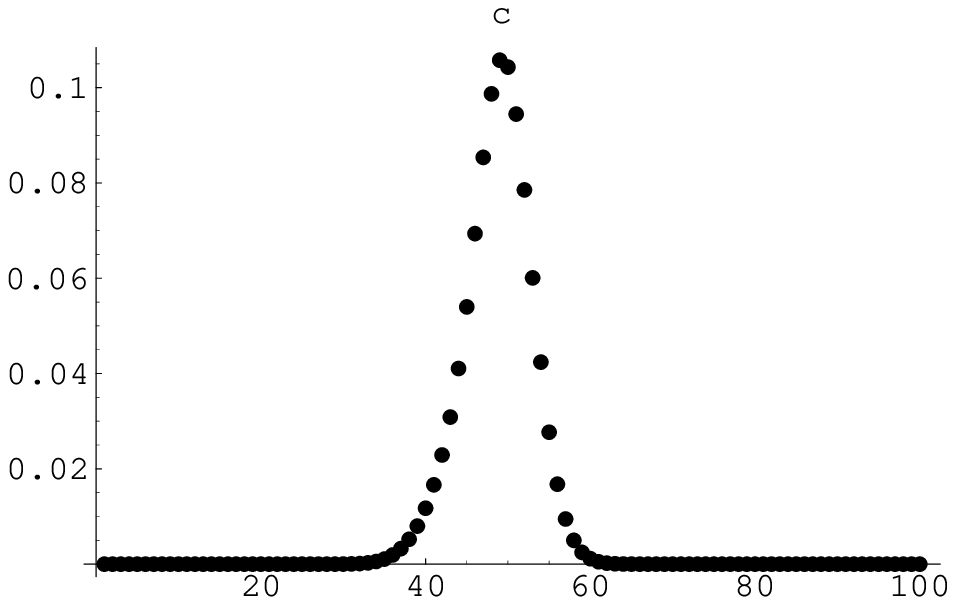}}
\mbox{\includegraphics[width=2.50in]{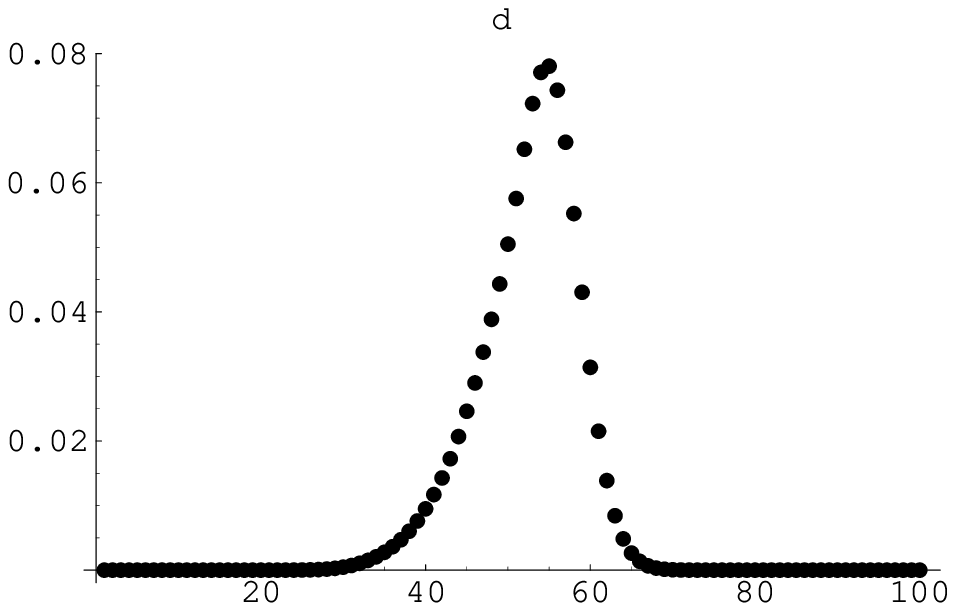}}
}
\vspace*{0.5in}
\centerline{
\mbox{\includegraphics[width=2.50in]{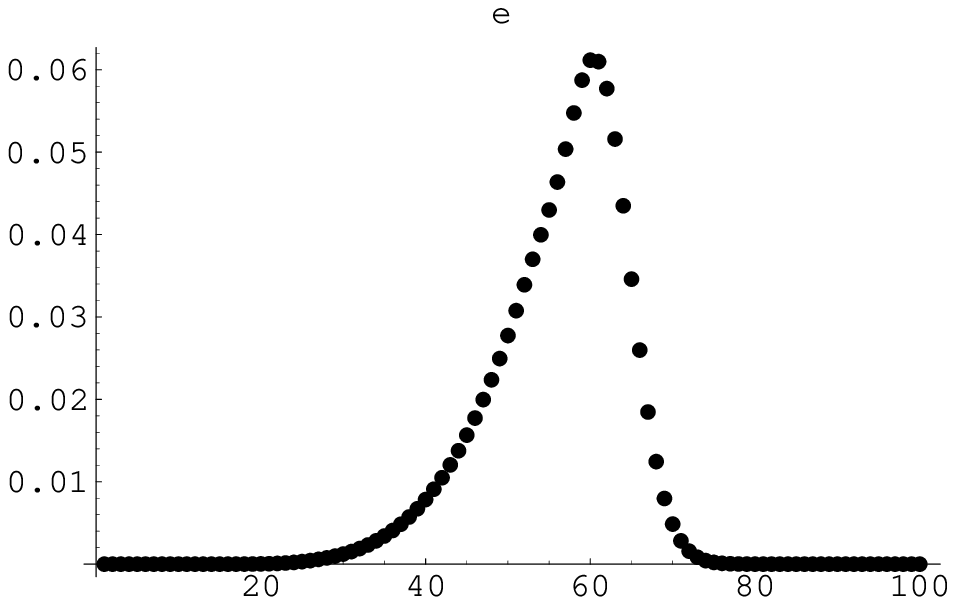}}
\mbox{\includegraphics[width=2.50in]{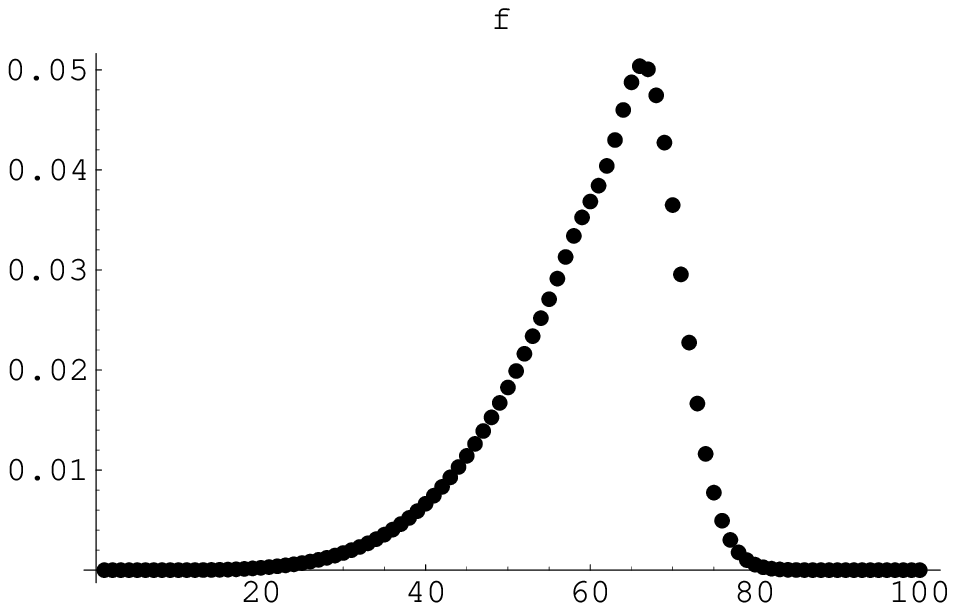}}
}
        \caption{Evolution of a MUS wave packet for $N = 100,k=2$ with 
$\mu=1.5, \lambda=-1.497+0.094i$. The plots correspond to times 
(a) $t=0$, (b) $t=50\pi/100$, (c) $t=100\pi/100$, (d) $t=150\pi/100$,
(e) $t=200\pi/100$, and (f) $t=250\pi/100$. The motion is, for small times,
the same as the spreading of a Gaussian wave packet on the line. The effects of 
the discretization become apparent at larger times.}
        \label{fig:k100c}
\end{figure}

Examples of time evolution are shown in Fig.~\ref{fig:k2a}
where $N=8$, $k=2$, in Fig.~\ref{fig:k4a} where
$N=8$, $k=4$, in Figs.~\ref{fig:k100a}-\ref{fig:k100b}
 where $N=100$, $k=25$, and in
Fig.~\ref{fig:k100c} where $N=100$, $k=2$.
The points correspond to the probability distribution $|c_j|^2$, and the
wave packet at $t=0$ in Figs.~\ref{fig:k2a}-\ref{fig:k4a}
is taken to be the minimum uncertainty state shown in 
Fig.~\ref{fig:cj}. The demonstration that the motion is periodic, and thus 
that a revival occurs, proceeds by brute force 
examination of all the terms in the time evolution operator. 
It would be 
desirable to obtain a more elegant proof of the periodicity. For the 
special cases $N/k=2,4,6$ one can sum the time evolution into trigonometric 
functions, in which case, the periodicity becomes obvious. We find 
for an arbitrary ket $|u_j\rangle$ that
\bea
&&N/k=2: \qquad \exp \left [-i\hat{H}t\right ]|u_j\rangle =\cos (t) |u_j\rangle
+\sum_{i\ne j}c_i(t)|u_i\rangle\;,
\nonumber \\
&&N/k=4: \qquad \exp \left [-i\hat{H}t\right ]|u_j\rangle =\cos^2 (t) |u_j\rangle+\sum_{i\ne j}c_i(t)|u_i\rangle\;,
\nonumber \\
&&N/k=6: \qquad \exp \left [-i\hat{H}t\right ]|u_j\rangle =\left ({2\over 3}\cos^2 (t) 
+{2\over 3}\cos(t)-{1\over 3}\right )|u_j\rangle+\sum_{i\ne j}c_i(t)|u_i\rangle\;.
\label{smallN}
\eea
Since the time evolution operator is unitary, when the square of the coefficient of $|u_j\rangle$, namely $|c_j|^2$, returns to one, a revival must occur.
This exactly periodic behavior will occur for any state, not necessarily the 
MUS. The periods are $\pi/2$, $\pi$ and $2\pi$ in our dimensionless time. 
We have also demonstrated numerically that there is a revival of the $t=0$ wave
packet at later times for higher values of $N/k$ for the MUS only. This
behavior seems to be generic, while the periodic behavior for the smaller 
values of $N/k$ in Eq.~(\ref{smallN}) are special cases.

The case of most interest to us is $N\gg 1$ and $N/k\gg 1$, 
which corresponds to 
the case where the wave packet is close to the continuum Gaussian case and 
the motion for small times is the evolution of this Gaussian according to 
a free particle Hamiltoian. For this case our numerical method is unable to 
follow the time evolution until a revival occurs. A rigorous demonstration that
the revivals that occur for small $N/k$ also occur for these very 
large $N/k$ is still 
needed. 

When one takes the limit in which the phase space radii to 
infinity so that that discretized phase space approaches the continuum
(and $N$ goes to infinity), 
there is a residual part of the Hamiltonian that does not represent a 
free particle on the line (for which we know that there is a spreading 
of the wave packet). The spreading of the wave packet on the line does not 
represent a violation of Liouville's theorem which states the volume of 
phase space under unitary time evolution must be invariant. In fact it can be
easily shown that if one uses the quantum operators $\hat p$ and 
$\hat{x}'=\hat{x}-\hat{p}/mt$, the 
dispersions of a Gaussian wave packet saturates the uncertainty principle for 
all time (In the standard variables $p$ and $x$, the spreading of the wave 
packet causing the product of the dispersions, $(\Delta \hat{x})^2(\Delta \hat{p})^2$,
to increase from the 
minimum uncertainty wave packet. Of course under time reversal the product 
of the dispersions can be made to decrease until it reaches a minimum of $1/4$,
after which it will subsequently increase. There is no contradiction with 
Liouville's theorem as a rectangle will evolve into a parallelogram under time 
evolution and preserve the area. It is easy to see that the operator $x'$ 
represents the position coordinate with the average ``velocity'' of the wave
packet taken into account.) The periodic behavior exhibited by the wave 
packets in the discrete quantum phase space can be seen to arise from the 
presence of these small correction factors which accumulate when times 
become large enough for the particle to make its way around the compactified 
direction.

\section{Generalized Uncertainty Principle}

For topologically nontrivial configuration space it is known that there are
consequences for the uncertainty principle. The simplest illustration is the 
motion of a particle on a circle. The appropriate position 
operator to employ is the phase operator rather than the angle operator which 
is not single-valued.  The (angular) momentum is then quantized unlike the 
usual quantum mechanics of a particle on a line. The wave function can be in 
a definite state of angular momentum ($\psi\sim \exp (i\langle L\rangle \phi)$)
in which case the probability distribution is constant on the circle.
Clearly if one insists on using the dispersions $\Delta L$ and $\Delta \phi$
for an uncertainty principle, then (since the state is an $L$-eigenstate) 
the dispersion in $\Delta L$ is zero. 

In this section we will be interested in showing that there are small 
corrections to the usual uncertainty principle in position $Q$ and momentum $P$
when the continuum phase space is approximated by a discrete one.
It is expected that a discretization of space or the existence of a 
minimum length will result in modifications to the uncertainty principle. In
particular one expects corrections of the form
\begin{eqnarray}
\Delta x \ge {{1}\over {2\Delta p}}+ \alpha \ell_P^2 \Delta p + \dots
\;.
\label{gup}
\end{eqnarray}
One obtains this expression along with the coefficient $\alpha $ in the case
momentum is compactified on a circle. Actually one expects a whole series of
terms on the right hand side involving higher order quantities such as 
$\langle p^4 \rangle$. For the toy model of momentum compactified on a circle,
the full expression can be worked out in detail\cite{Bang:2006va}, and a 
smooth extrapolation to the limit where the discretization of space becomes 
dominant can be performed. 
The uncertainty priniciple is easily generalized from the case involving
Hermition operators to the case of unitary operators $\hat{U}$ and $\hat{V}$, and has the 
following form
\bea
\left\langle \Delta \hat{V}^\dag\Delta \hat{V}\right\rangle\left\langle \Delta \hat{U}^\dag\Delta \hat{U}\right\rangle &=&\left|\left\langle \Delta \hat{V}^\dag\Delta \hat{U} \right\rangle \right|^2.\label{GUP2}\;
\eea
For the case where operators are Hermitian, this reduces to the well-known
form. The right hand side can be written
\bea
\left|\left\langle \Delta \hat{V}^\dag\Delta \hat{U} \right\rangle \right|^2&=&\left|\left<\hat V^\dag\hat U \right>-\left<\hat V^\dag \right>\left<\hat U \right>\right|^2\\
&=&\left|\alpha\beta\left<\hat P\hat Q\right> -{i\alpha\beta^2\over{2}\left<\hat P^2\hat Q \right>}-{\alpha\beta^3\over 6}\left<\hat P^3\hat Q \right>+{i\alpha^2\beta\over 2}\left<\hat P\hat Q^2 \right>\right.\nonumber\\
&&\left.+{\alpha^2\beta^2\over 4}\left(\left<\hat P^2\hat Q^2 \right>-\Delta\hat P^2\Delta\hat Q^2 \right)-{i\alpha^2\beta^3\over{12}}\left(\left<\hat P^3\hat Q^2 \right>-\left<\hat P^3 \right>\left<\hat Q^2 \right> \right)+\ldots \right|^2\;
\eea
while the left hand side gives
\bea
\left\langle \Delta \hat{V}^\dag\Delta \hat{V}\right\rangle&=&1-\left\langle \hat{V}\right\rangle\left\langle \hat{V}^\dag\right\rangle\\
&\approx&\beta^2\Delta \hat P^2\left[1-\beta^2\left({\Delta\hat{P}^2\over4}+{\left<\hat P^4\right>\over{}12\Delta \hat P^2}\right)+\ldots\right], \\
\left\langle \Delta \hat{U}^\dag\Delta \hat{U}\right\rangle&=&1-\left\langle \hat{U}\right\rangle\left\langle \hat{U}^\dag\right\rangle \\
&\approx&\alpha^2\Delta \hat Q^2\left[1-\alpha^2\left({\Delta\hat{Q}^2\over4}+{\left<\hat Q^4\right>\over{}12\Delta \hat Q^2}\right)+\ldots\right], \;
\eea
where $\Delta\hat{Q}^2=\left\langle \hat{Q}^2\right\rangle-\left\langle \hat{Q}\right\rangle^2$ and (without loss of generality) we take $\left\langle \hat{Q}\right\rangle=0,$ and the similarly for $\Delta\hat{P}^2.$
Then we have
\bea
\left\langle \Delta \hat{V}^\dag\Delta \hat{V}\right\rangle\left\langle \Delta \hat{U}^\dag\Delta \hat{U}\right\rangle&=&\alpha^2\beta^2\Delta\hat{Q}^2\Delta\hat{P}^2\left[ 1-{1\over 4}\left(\alpha^2\Delta\hat{Q}^2+\beta^2\Delta\hat{P}^2 \right)\right.\nonumber\\
&&+\;\left.{\alpha^2\beta^2\over{16}}\Delta\hat Q^2\Delta\hat P^2+\ldots\right]
\eea
Keeping only the leading term, one obtains
\bea
\Delta\hat{Q}^2\Delta\hat{P}^2&\approx &\left|\left<\hat P\hat Q\right>\right|^2 =\left|{\hbar}\over{2i} \right|^2={\hbar^2\over 4}\;,
\eea
where $\left|\left<\hat P\hat Q\right>\right|^2$ is calculated for a gaussian wave packet on the line:
\bea
\psi(Q)&=&\left[2\pi \Delta Q^2\right]^{-1/4}\exp\left[-\left({{Q-\left<Q\right>}\over{2\Delta Q}}\right)^2 \right],\\
\left<\hat P\hat Q \right>&=&{{\hbar /i}\over{\sqrt{2\pi\Delta Q^2}}}\int_{-\infty}^{\infty}dQ\psi^*{d\over{dQ}}\left[Q*\psi(Q)\right]\nonumber\\
&=&{{\hbar /i}\over{\sqrt{2\pi\Delta Q^2}}}\int_{-\infty}^{\infty}dQ\left\{{{\left<Q \right>Q-Q^2+2\Delta Q^2 }\over{2\Delta Q^2 }} \right\}\exp\left[-{{\left(Q-\left<Q \right> \right)^2 }\over{2\Delta Q^2 }} \right]\nonumber\\
&=&{{\hbar /i}\over{\sqrt{2\pi\Delta Q^2}}}\left\{\sqrt{{\pi \over 2}}\Delta Q \right\}\nonumber\\
&=&\hbar /2i\;.
\eea
This demonstrates that the uncertainty relation for the minimum uncertainty 
wave packet approaches the minimum uncertainty Gaussian in the continuum 
limit. Keeping the first subleading terms one obtains
\bea
\Delta\hat{Q}^2\Delta\hat{P}^2&\approx&{\hbar^2\over 4}\left \{1+ {1\over N^{1/2}}\left({\ell_p^2\pi\over{2\hbar^2}}\Delta\hat P^2\right )
+{1\over N^{3/2}}\left({\pi\over{2\ell_p^2}}\Delta\hat Q^2\right )+\ldots  \right \}\;.
\eea

The coefficients on the right hand side reflect the choice for the parameters
$\alpha $ and $\beta$ in Eq.~(\ref{defnalphabeta}).
The uncertainty principle in $\hat{U}$ and $\hat{V}$ is 
understood as a generalized 
uncertainty principle when interpreted in terms of the approximate Hermitian 
operators $\hat{Q}$ and $\hat{P}$ of the continuum.

\section{Conclusions}

The goal of unifying quantum mechanics and gravity suggests that the underlying
spacetime is in fact discrete at the Planck scale. This discretization is not 
unique, and it is not clear even what the fundamental variables one should use 
are. In order to study the possible effects of such discretization we have 
looked at a one-dimensional quantum mechanical model that is completely 
discrete (in configuration and momentum space), and have studied the 
modifications that result to the continuum case. These modifications include
extra terms in the uncertainty principle that are suppressed by the 
Planck length. 
That such corrections may generally be present in theories of 
quantum gravity (such as string theory) has been known for some time.

For fine discretizations the minimum uncertainty state will 
appear to spread in the usual fashion of the Gaussian wave packet on the line, 
but after a sufficient time it will arrive back at its original position with 
approximately the 
same coefficients $c_j$. This is a new form of revival of wave packets that
does not require an external potential, but arises from the topology
of the phase space. For small values of $N/k \le 6$ 
the periodic behavior is exact
and is exhibited by all states. For larger values of $N/k$ the revivals 
are approximate and only occur for the minimum uncertainty states.
In particular the minimum uncertainty state which approximates the 
Gaussian wave packet of a quantum particle in the continuum limit
spreads in the usual way under time evolution but is expected at large times
to exhibit a revival.

\section*{Acknowledgments}
%\begin{theacknowledgments}
This work was supported in part by the U.S.
Department of Energy under Grant No.~DE-FG02-91ER40661.
%\end{theacknowledgments}

\end{document}